\documentclass[conference]{IEEEtran}
\IEEEoverridecommandlockouts
\usepackage{cite}
\usepackage{amsmath,amssymb,amsfonts}
\usepackage{algorithmic}
\usepackage{graphicx}
\usepackage{textcomp}
\usepackage{xcolor}
\usepackage{booktabs, multirow,array}
\usepackage{float}
\usepackage{subfig}
\usepackage{upgreek}

\def\BibTeX{{\rm B\kern-.05em{\sc i\kern-.025em b}\kern-.08em
    T\kern-.1667em\lower.7ex\hbox{E}\kern-.125emX}}
\begin{document}

\title{Diagonal Waveform and Algorithm to Estimate Range and Velocity in Multi-Object Scenarios}

\author{\IEEEauthorblockN{Yi Geng}
	\IEEEauthorblockA{\textit{Cictmobile, China} \\
		gengyi@cictmobile.com}
}

\maketitle

\begin{abstract}
Waveform design for joint communication and sensing (JCAS) is an important research direction, focusing on providing an optimal tradeoff between communication and sensing performance. In this paper, we first describe the conventional grid-type waveform structure and the corresponding two-dimension (2D)-discrete Fourier transform (DFT) algorithm. We then introduce an emerging diagonal scheme, including a diagonal waveform structure and corresponding 1D-DFT diagonal algorithm. The diagonal scheme substantially reduces the signaling overhead and computational complexity compared to the conventional \mbox{2D-DFT} algorithm while still achieving the same radar performance. But the previous study of diagonal waveform used a single target to evaluate the performance of the diagonal scheme. This paper verifies the diagonal waveform with simulations demonstrating its feasibility in a traffic monitoring scenario with multiple vehicles.
\end{abstract}


\section{Introduction}
Joint communication and sensing (JCAS) is an emerging technology in 6G that localize and track passive targets or extract characteristics from targets (e.g., material information\cite{9482524}) using wireless communication infrastructures. JCAS system needs to guarantee an appropriate radar performance by using the time-frequency resource shared between the communication and sensing functionalities. From a waveform design perspective, optimization of sensing signals, such as code sequence selection, modulation scheme, precoding, and power allocation, may conflict with the requirements of communication signals \cite{9839260}. Therefore, dedicated sensing signals are desired. Waveform design for the JCAS system is formulated as an optimization problem to find a proper tradeoff between sensing and communication. For a fixed time-frequency resource block, the communication rate can be increased by degrading the radar performance and vice versa\cite{9737357}.

\begin{figure}[t]
	\centerline{\includegraphics[width=\linewidth, height=10cm, keepaspectratio]{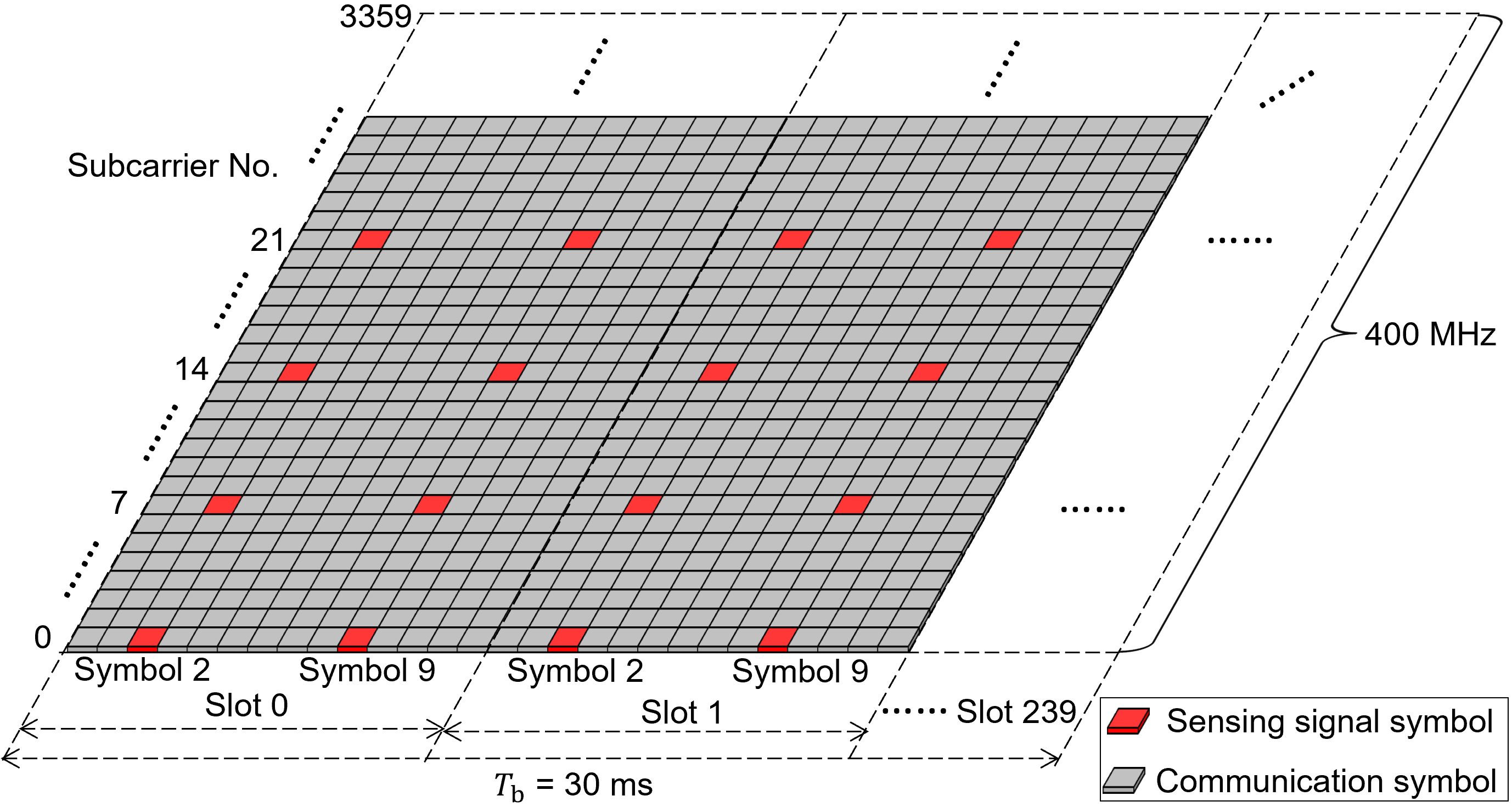}}
	\caption{A grid structure of sensing signals.}
	\label{fig_1}
\end{figure}

The sensing signals of orthogonal frequency division multiplexing (OFDM)-based JCAS systems can be allocated sparsely in the frequency and/or time domains to reduce sensing overhead, resulting in grid structures or comb structures. For example, an exemplary waveform in the grid structure is emitted for road traffic monitoring, as illustrated in Fig.~1. The transmitter of an OFDM-based JCAS system sends a time-frequency resource block with a bandwidth of 400~MHz and a duration of 30~ms consecutively. All the OFDM parameters used in this paper are listed in Table~I. In the frequency domain, the sensing signals are transmitted every seven subcarriers. In the time domain, the sensing signals are transmitted every seven symbols. Therefore, there are 480 sensing signals (colored in red in Fig.~1) both in frequency and time domain in one resource block. The resources colored in gray are allocated for communication. According to the parameter dependencies of OFDM radar \cite{9354629}, the waveform in Fig.~1 would be sufficient to support a range resolution of 0.4~m, a velocity of 0.2~m/s, a maximum unambiguous detection range of 179~m, and a maximum unambiguous detection velocity of 92~m/s, which can meet the requirements of traffic monitoring applications. Assuming that the sensing blocks are transmitted consecutively, the sensing signal overhead of this waveform is 2\%, which corresponds to the overhead for sensing in one beam direction. Considering that highly directional beams are needed to compensate for the high path loss at 28~GHz, the sensing beams must be emitted in many directions. Thus the sensing overhead increases further \cite{9743513}. As a result, waveform structures with low overhead need to be developed.

The paper is structured as follows. Section~II presents the conventional modulation-symbol-based OFDM radar processing algorithm. Section~III presents an emerging diagonal waveform with low overhead and verify it in a single-target scenario. In Section~IV, we extend the scheme in Section~III to scenarios with multiple reflecting targets. Finally, the conclusions are drawn in Section~V.
\begin{table}[t]
	\caption{OFDM parameters for 2D-DFT scheme and diagonal scheme\label{tab:table2}}
	\begin{center}
		\begin{tabular}{ccc}
			\hline
		Parameter & Symbol & Value\\
\hline
Carrier frequency & $f_\text{c}$ & 28 GHz\\
Total signal bandwidth & $B$ & 400~MHz\\
Subcarrier spacing & $\Delta{f}$ & 120 KHz\\
OFDM symbol duration & $T_{\text{sym}}$ & 8.92 $\upmu{s}$\\
Block duration & $T_\text{b}$ & 30~ms\\
Subcarriers in $B$ & $N_\text{c}$ & 3360\\
Symbols in $T_\text{b}$ & $N_\text{{sym}}$ & 3360\\
Sensing signals in $B$ & $N_\text{f}$ & 480\\
Sensing signals in $T_\text{b}$ & $N_\text{t}$ & 480\\
Sensing signals in diagonal & $N$ & 480\\
\hline
		\end{tabular}
	\end{center}
	\label{tab_2}
\end{table}

\section{Modulation-Symbol-Based 2D-DFT Algorithm}
One of the conventional algorithms for grid-type sensing signals, based on modulation-symbol, is employed to estimate the range and velocity of the reflecting objects \cite{comb}. The conventional algorithm is summarized in this section. The presence of reflecting objects in the detection range affects the propagation of radio waves in the air. The grid-type received sensing signal in the modulation symbol domain $\mathbf{c}_{\text{Rx}}(m,n)$, which consists of a superposition of the range and velocity information of the reflecting objects upon the transmitted sensing signal $\mathbf{c}_{\text{Tx}}(m,n)$ in the modulation symbol domain, is given by
\begin{equation}\label{eqn_1}
	\mathbf{c}_{\text{Rx}}(m,n)=A(m,n)\mathbf{c}_{\text{Tx}}(m,n)\phi_\text{R}(m)\otimes{\phi_\text{D}(n)},
\end{equation}
where $\phi_\text{R}(m)$ is a $N_\text{f}\times{1}$ vector, $\phi_\text{D}(n)$ is a $1\times{N_\text{t}}$ vector, and
\begin{equation}\label{eqn_2}
	\phi_\text{R}(m) = \text{exp}({\frac{-\text{j}4\pi\Delta{f}Rm}{\text{c}}}), m=0, \cdots,N_\text{f}-1
\end{equation}
\begin{equation}\label{eqn_3}
	\phi_\text{D}(n) = \text{exp}({\frac{\text{j}4\pi{T_{\text{sym}}}{f_{\text{c}}}vn}{\text{c}}}), n=0, \cdots,N_\text{t}-1
\end{equation}
where $A(m,n)$ represents the attenuation and phase shift due to the propagation and reflection, $m$ is the individual subcarrier index from a total number of $N_\text{f}$ subcarriers, $n$ represents the individual OFDM symbol index within the total number of $N_\text{t}$ symbols, $\otimes$ referring to dyadic product, $R$ is the range between the JCAS node and the reflecting object, $v$ is the radial velocity of the reflecting object.

The normalized sensing signal matrix, $\mathbf{C}(m,n)$, can be obtained by element-wise division between $\mathbf{c}_{\text{Tx}}(m,n)$ and $\mathbf{c}_{\text{Rx}}(m,n)$, yielding a $N_\text{f}\times{N_\text{t}}$ matrix
\begin{equation}\label{eqn_4}
	\mathbf{C}(m,n)=\frac{\mathbf{c}_{\text{Rx}}(m,n)}{\mathbf{c}_{\text{Tx}}(m,n)} = A(m,n)\phi_\text{R}(m)\otimes{\phi_\text{D}(n)}.
\end{equation}

The linear phase shifts of the columns (or rows) of matrix $\mathbf{C}(m,n)$ contain the range (or Doppler) information of the target. Applying IDFT to vector $\phi_\text{R}(m)$, yielding
\begin{multline}\label{eqn_5}
	\Phi_r(p) = \text{IDFT}(\phi_\text{R}(m))=\frac{1}{N_\text{f}}\sum_{m=0}^{N_\text{f}-1}\phi_\text{R}(m)\text{exp}({\frac{\text{j}2\pi{m}p}{N_\text{f}}}),\\
	p=0, \cdots,N_\text{f}-1
\end{multline}

The IDFT response $\Phi_r(p)$ shows a peak at IDFT bin index $p_{\text{peak}}$. Then the range $R$ can be calculated by
\begin{equation}\label{eqn_6}
	R = \frac{\text{c}p_{\text{peak}}}{2\Delta{f}N_\text{f}}.
\end{equation}

Similarly, applying DFT to vector $\phi_\text{D}(n)$, yielding
\begin{multline}\label{eqn_7}
	\Phi_v(q) = \text{DFT}(\phi_\text{D}(n))=\sum_{n=0}^{N_\text{t}-1}\phi_\text{D}(n)\text{exp}({-\frac{\text{j}2\pi{n}q}{N_\text{t}}}),\\
	q=0, \cdots,N_\text{t}-1
\end{multline}

The DFT response $\Phi_v(q)$ shows a peak at DFT bin index $q_{\text{peak}}$. Then the velocity $v$ can be calculated by
\begin{equation}\label{eqn_8}
	v = \frac{\text{c}q_{\text{peak}}}{2f_cT_{\text{sym}}N_\text{t}}.
\end{equation}

In practice, 2D-DFT of matrix $\mathbf{C}(m,n)$ is performed to derive the range and velocity of the reflecting object simultaneously. 2D-DFT of $\mathbf{C}(m,n)$ is equivalent to column-by-column 1D-IDFTs of length $N_\text{f}$ after row-by-row 1D-DFTs of length $N_\text{t}$.
\addtolength{\topmargin}{0.03in}
\section{A Diagonal Waveform Design and Corresponding Algorithm}
From the implementation point of view, the high computational complexity of large-scale DFT/IDFT calculations creates a bottleneck and limits the sensing performance of the 2D-DFT algorithm \cite{6026390}. To derive a range-velocity measurement from a $N\times{N}$ modulation symbol matrix $\mathbf{C}(m,n)$, $2N\mathcal{O}(N^2)$ complex multiplications are needed, including $N$ DFTs on $N$ rows and $N$ IDFTs on $N$ columns. In addition, the grid structure incurs a large sensing signal overhead and degraded communication performance.

To tackle these problems, we have proposed a diagonal waveform structure, and corresponding diagonal algorithm for single-target scenario in our previous work \cite{10107516}. The scheme in \cite{10107516} is as follows. The JCAS system uniformly allocates the sensing signals along the resource block diagonal and reserves the resources outside the diagonal for communication. This structure guarantees that the same number of sensing signals, $N$, is transmitted both in frequency band $B$ and in time duration $T_\text{b}$. As depicted in Fig.~2(a), the transmitter emits one block of radio signals, including 480 sensing signals along the diagonal of the block and user data for communication. To ensure target tracking, the diagonal signals are emitted every 30~ms, as shown in Fig.~2(b). The normalized sensing signal vector, $\mathbf{d}(k)$, can be acquired by exploiting transmitted sensing signal vector $\mathbf{d}_{\text{Tx}}(k)$ and the received echos $\mathbf{d}_{\text{Rx}}(k)$, yielding
\begin{multline}\label{eqn_9}
	\mathbf{d}(k)=\frac{\mathbf{d}_{\text{Rx}}(k)}{\mathbf{d}_{\text{Tx}}(k)}\\=A(k)\text{exp}({\frac{-\text{j}4\pi\Delta{f}Rk}{\text{c}}})\text{exp}({\frac{\text{j}4\pi{T_{\text{sym}}}{f_{\text{c}}}vk}{\text{c}}}),\\k=0, \cdots,N-1
\end{multline}
where $A(k)$ represents the attenuation and phase shift due to the propagation and reflection, and
\begin{equation}\label{eqn_10}
	\phi_\text{R}(k) = \text{exp}({\frac{-\text{j}4\pi\Delta{f}Rk}{\text{c}}}),
\end{equation}
\begin{equation}\label{eqn_11}
	\phi_\text{D}(k) = \text{exp}({\frac{\text{j}4\pi{T_{\text{sym}}}{f_{\text{c}}}vk}{\text{c}}}).
\end{equation}

\begin{figure}[t]
	\centering
	\subfloat[A diagonal of sensing signals for one measurement]{\label{fig:5a}\includegraphics[width=0.9\columnwidth]{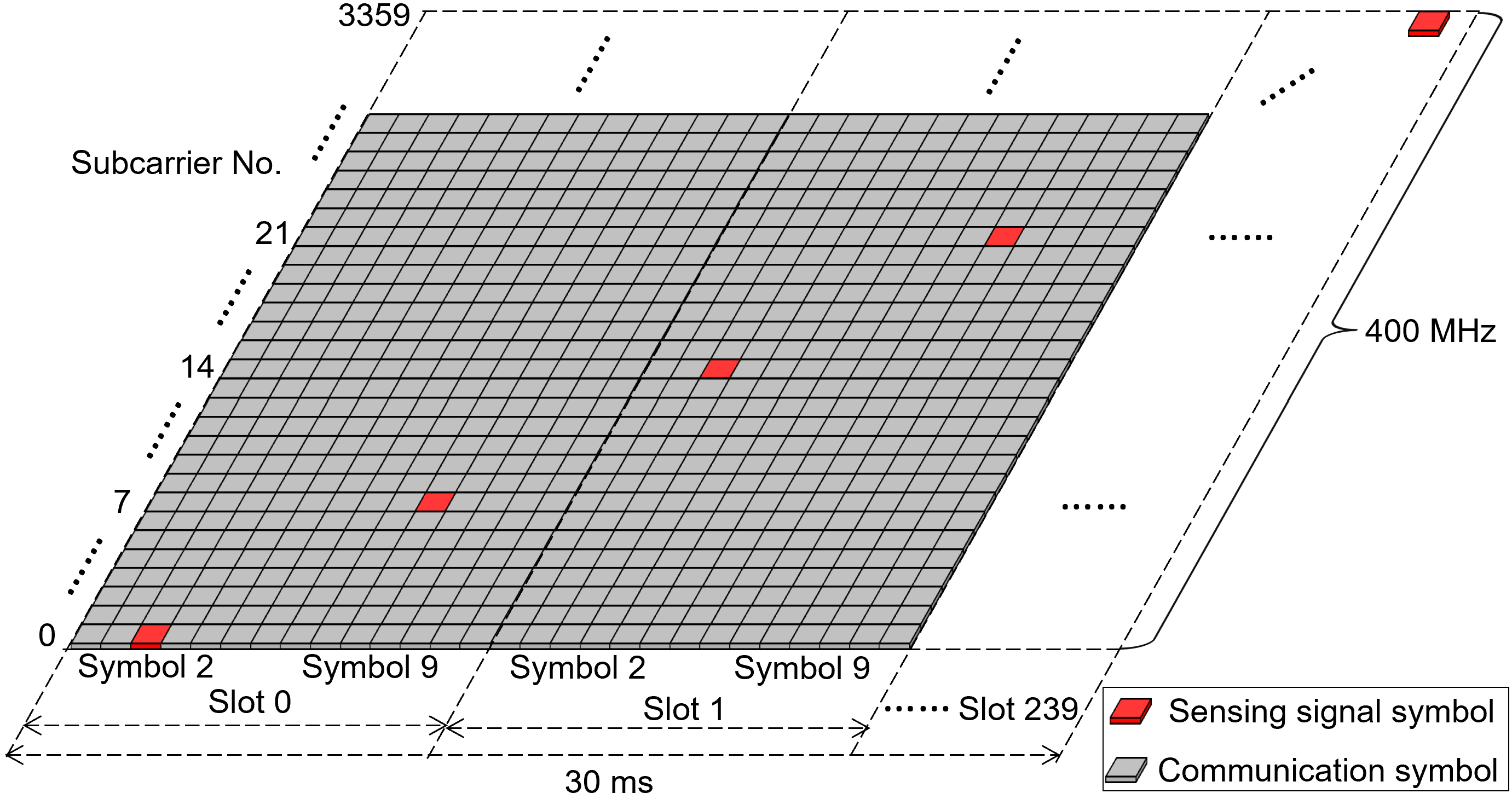}}\quad
	\subfloat[Consecutively transmitted diagonals of sensing signals]{\label{fig:5b}\includegraphics[width=1\columnwidth]{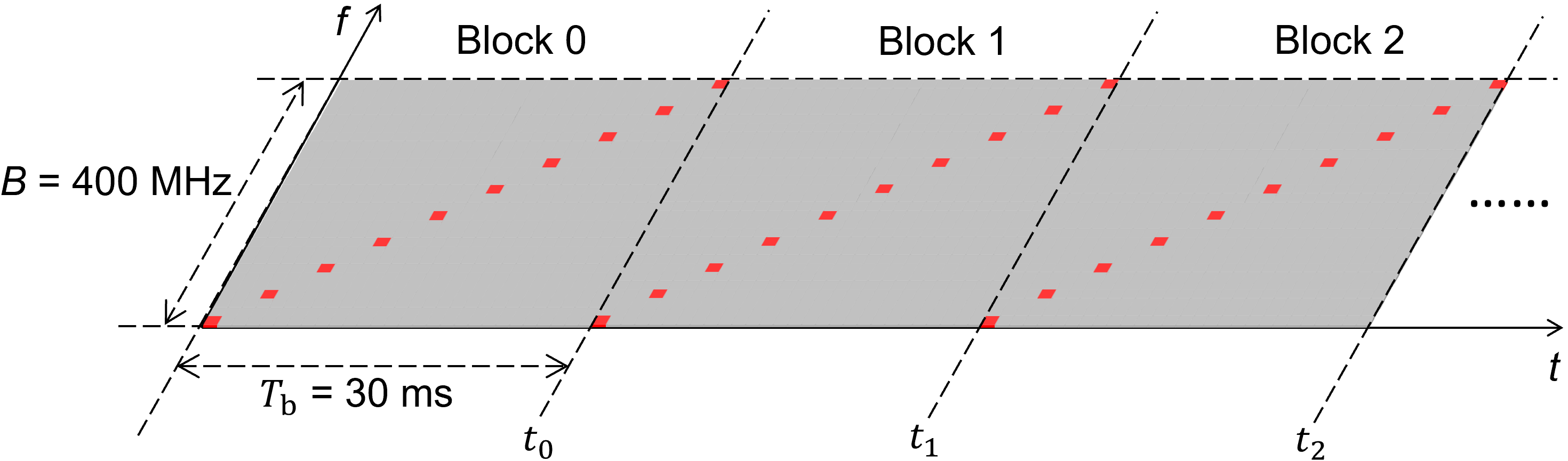}}\\
	\caption{The diagonal structure of sensing signals.}
	\label{fig_2}
\end{figure}

The modulation symbol vector $\mathbf{d}(k)$ simultaneously incorporates the phase shifts induced by the range $R$ and the velocity $v$ because the diagonal signals span both the frequency and time domains. Therefore, the range and velocity of the objects can be derived simultaneously by performing DFT to $\mathbf{d}(k)$, which yields
\begin{multline}\label{eqn_12}
	\Phi_{rv}(l)=\text{DFT}(\mathbf{d}(k))=
	\sum_{k=0}^{N-1}\phi_{\text{R}}(k)\phi_{\text{D}}(k)\text{exp}({-\frac{\text{j}2\pi{k}l}{N}}),\\
	l=0, \cdots,N-1
\end{multline}

The DFT of $\mathbf{d}(k)$ generates a dual-peak-like profile. Fig.~3 illustrates the dual-peak profile caused by an object with range \mbox{$R$ = 40~m} and velocity \mbox{$v$ = 5~m/s} by using the diagonal waveform and 1D-DFT diagonal algorithm. The dual-peak profile indicates two peaks at the DFT bin indices $l_1$ = 81 and $l_2$ = 134. A new term $l_\text{mean}$ is defined as the mean of the peak bin indices $l_1$ and $l_2$. Another term, $l_\Delta$, is defined as the difference between the peak bin indices $l_2$ and $l_1$. The terms $l_\text{mean}$ and $l_\Delta$ are given by
\begin{equation}\label{eqn_13}
	l_\text{mean} = \frac{l_1+l_2}{2},
\end{equation}
\begin{equation}\label{eqn_14}
	l_\Delta = l_2-l_1.
\end{equation}

One of $l_\text{mean}$ and $l_\Delta$ incorporates the range information, whereas the other incorporates the velocity information. The range $R$ and velocity $v$ of the object can be calculated by
\begin{equation}\label{eqn_15}
	R = \frac{\text{c}l_\text{mean}}{2\Delta{f}N_{\text{c}}}, v = \frac{
		\text{c}l_\Delta}{4T_{\text{sym}}f_{\text{c}}N_{\text{c}}},
\end{equation}
or
\begin{equation}\label{eqn_16}
	R = \frac{\text{c}l_\Delta}{4\Delta{f}N_{\text{c}}}, v = \frac{
	\text{c}l_\text{mean}}{2T_{\text{sym}}f_{\text{c}}N_{\text{c}}}.
\end{equation}

However, either \eqref{eqn_15} or \eqref{eqn_16} returns unexpected range-velocity values because of the unlabeled bin indices $l_1$ and $l_2$. We are uncertain which equation among \eqref{eqn_15} and \eqref{eqn_16} is indicative of the accurate range and velocity. For the case shown in Fig.~3, by using \eqref{eqn_15}, $R$ = 40~m and $v$ = 5~m/s can be derived. Or by using \eqref{eqn_16}, $R$ = 10~m and $v$ = 20~m/s is obtained. This means that a target with \mbox{$R$ = 40~m} and \mbox{$v$ = 5~m/s} induces the same dual-peak profile as that of another target with \mbox{$R$ = 10~m} and \mbox{$v$ = 20~m/s}. How to resolve this ambiguity has been discussed in \cite{10107516}.

\begin{figure}[t]
	\centerline{\includegraphics[width=\linewidth, height=6cm, keepaspectratio]{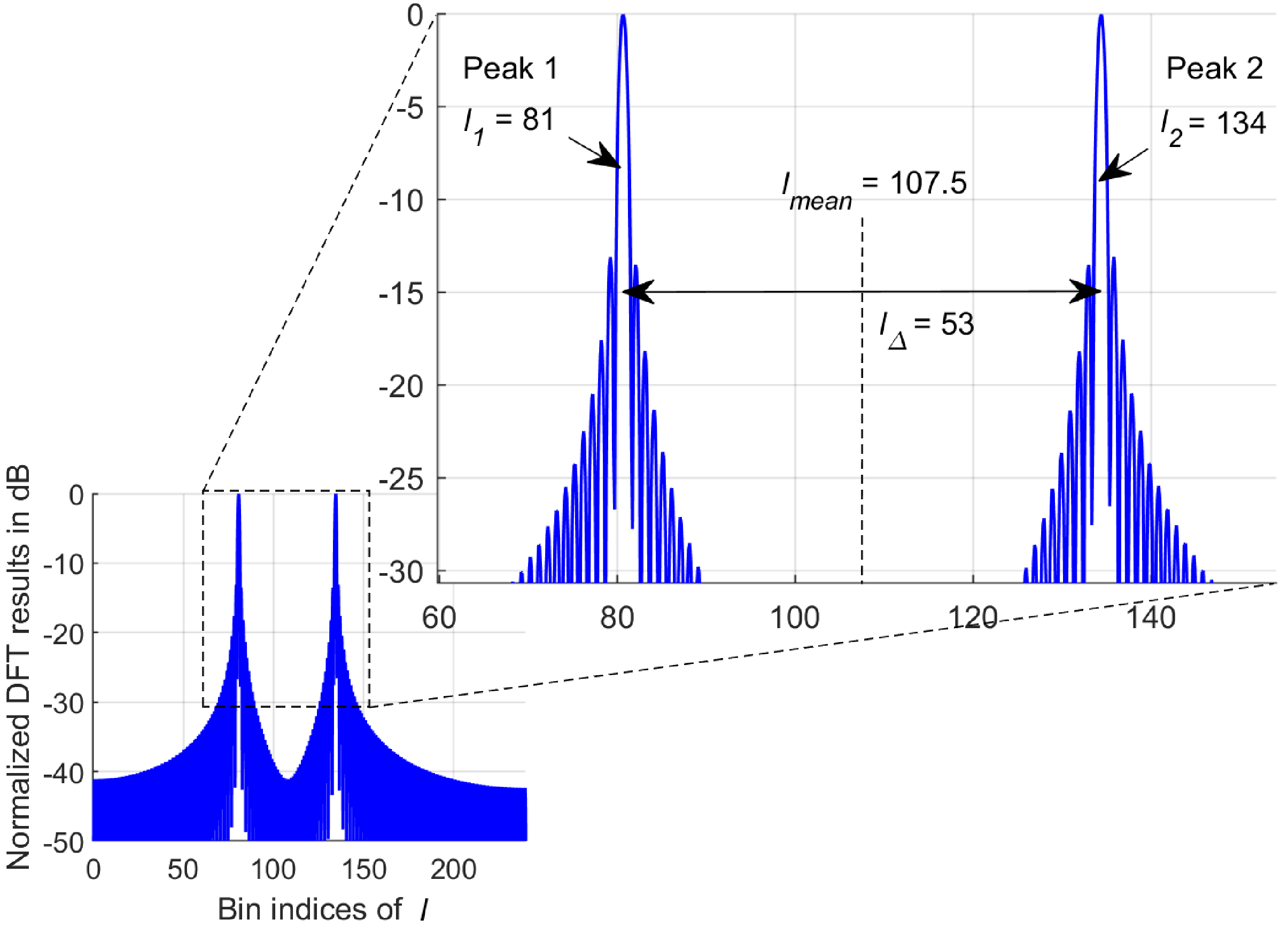}}
	\caption{The radar image of an object with range of 40~m and velocity of 5~m/s using the diagonal waveform and corresponding 1D-DFT diagonal algorithm.}
	\label{fig_3}
\end{figure}

\begin{figure*}[t]
	\centering
	\subfloat[Snapshot captured at time $t_0$ = 0~s, showing a yellow car traveling 6~m ahead on the left lane and a red car traveling 39~m ahead on the right lane]{\label{fig:h}\includegraphics[width=0.65\columnwidth]{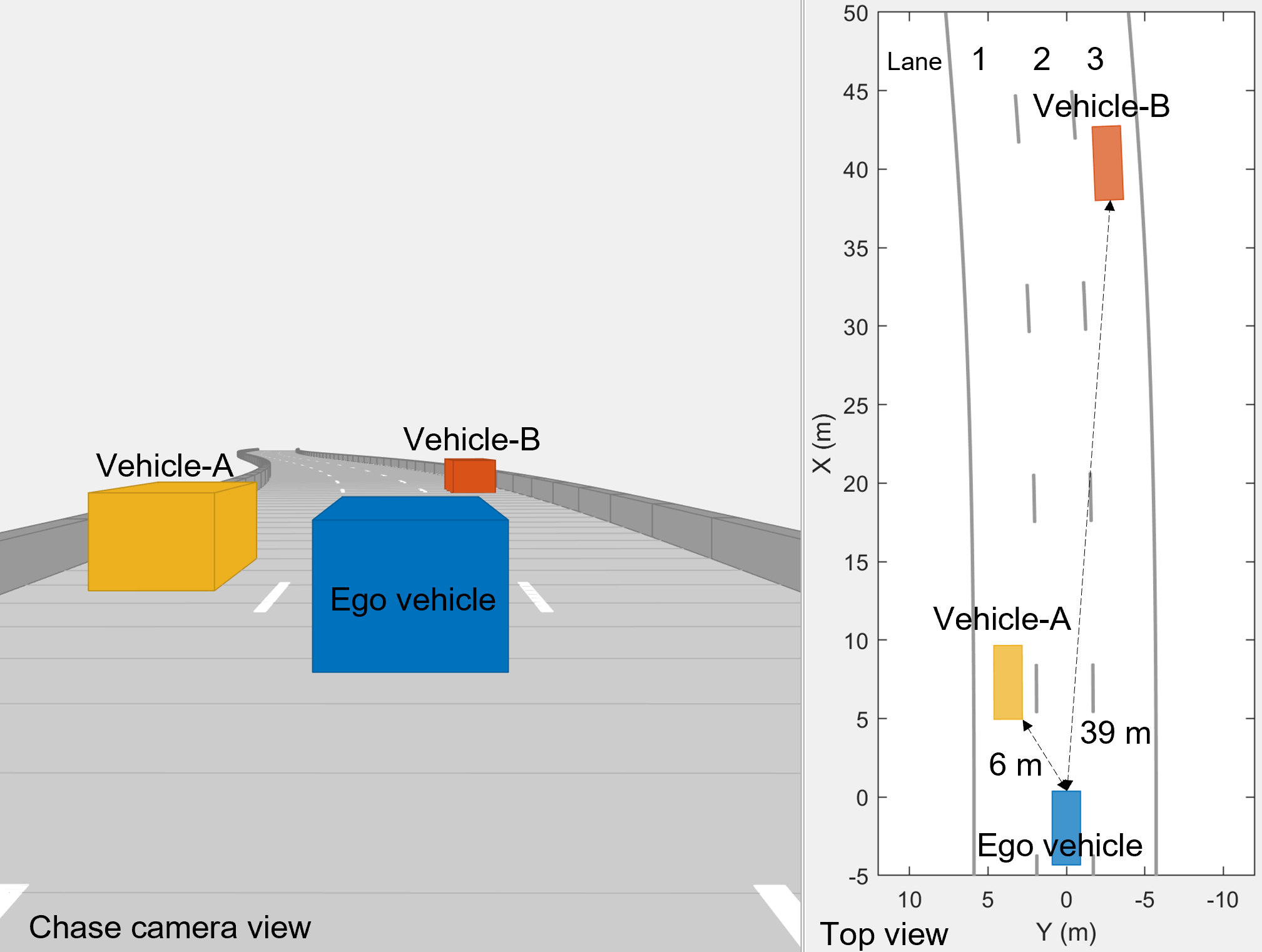}}\quad
	\subfloat[Snapshot captured at time $t_1$ = 0.2~s, showing a yellow car traveling 10~m ahead on the left lane and a red car traveling 40~m ahead on the right lane]{\label{fig:i}\includegraphics[width=0.65\columnwidth]{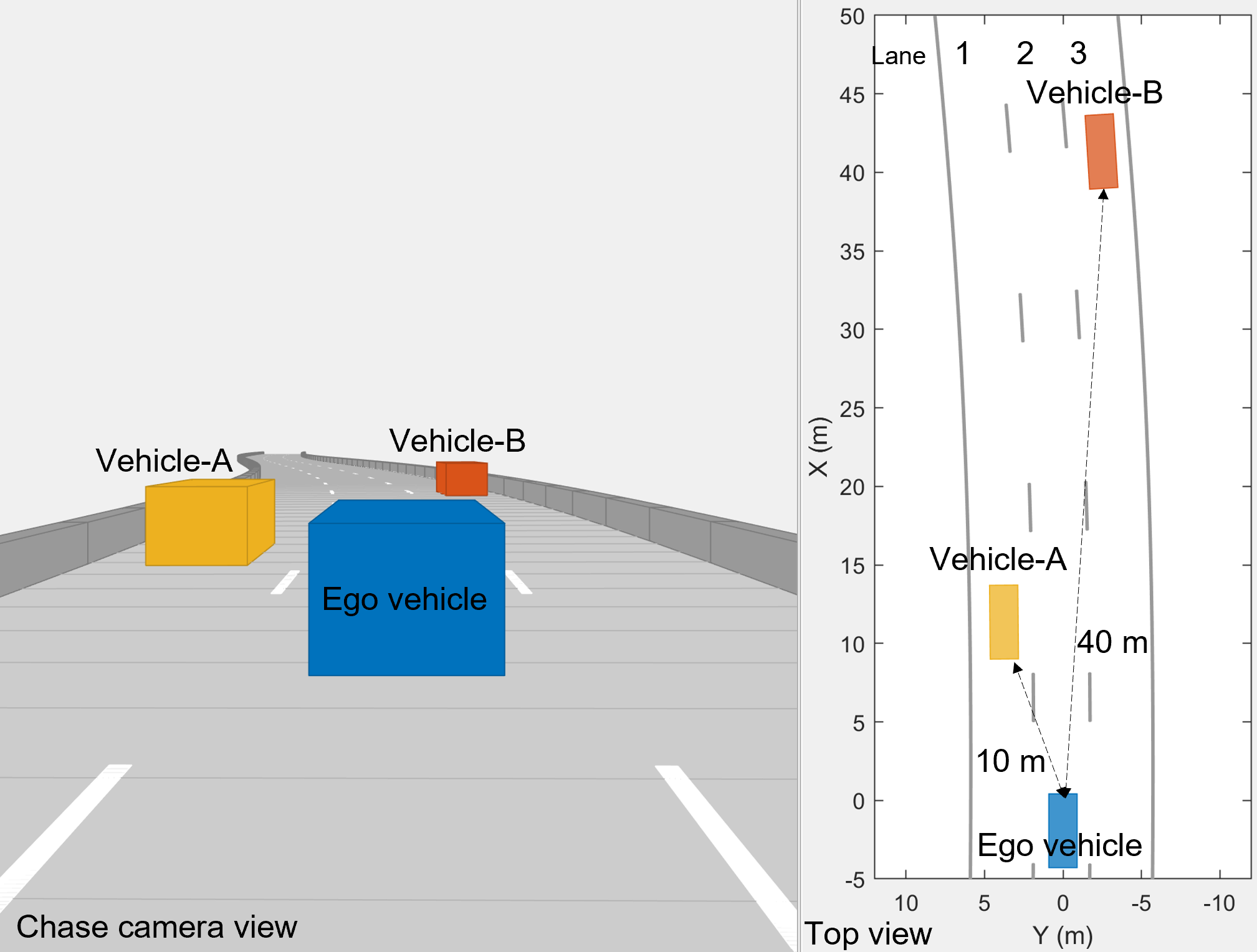}}\quad
	\subfloat[Snapshot captured at time $t_2$ = 0.6~s, showing a yellow car traveling 18~m ahead on the left lane and a red car traveling 42~m ahead on the right lane]{\label{fig:g}\includegraphics[width=0.65\columnwidth]{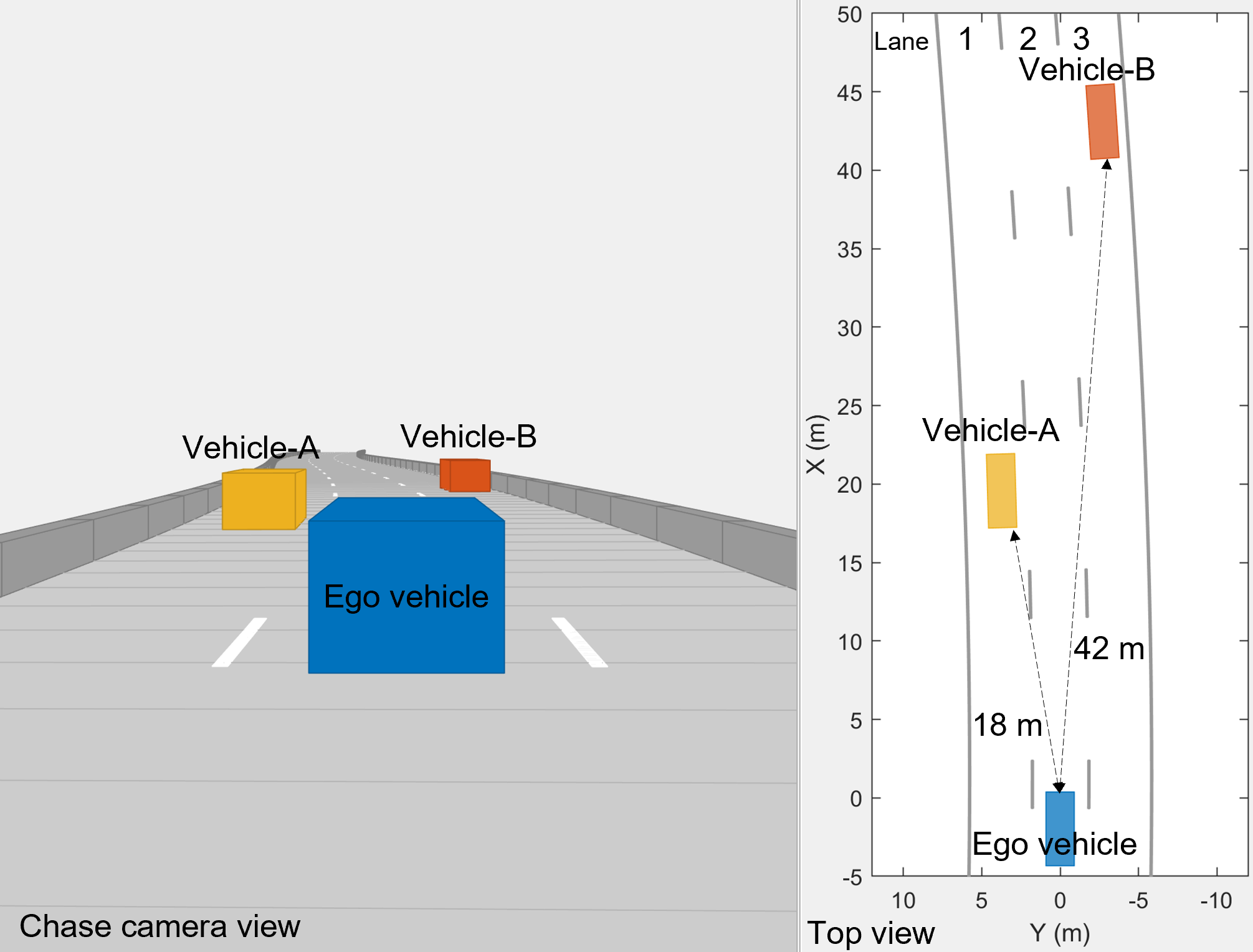}}\quad
	\subfloat[The radar image calculated at time $t_0$ = 0~s, vehicle-A with $R$ = 6~m and $v$ = 20~m/s, and vehicle-B with $R$ = 39~m and $v$ = 5~m/s , no windowing]{\label{fig:h}\includegraphics[width=0.65\columnwidth]{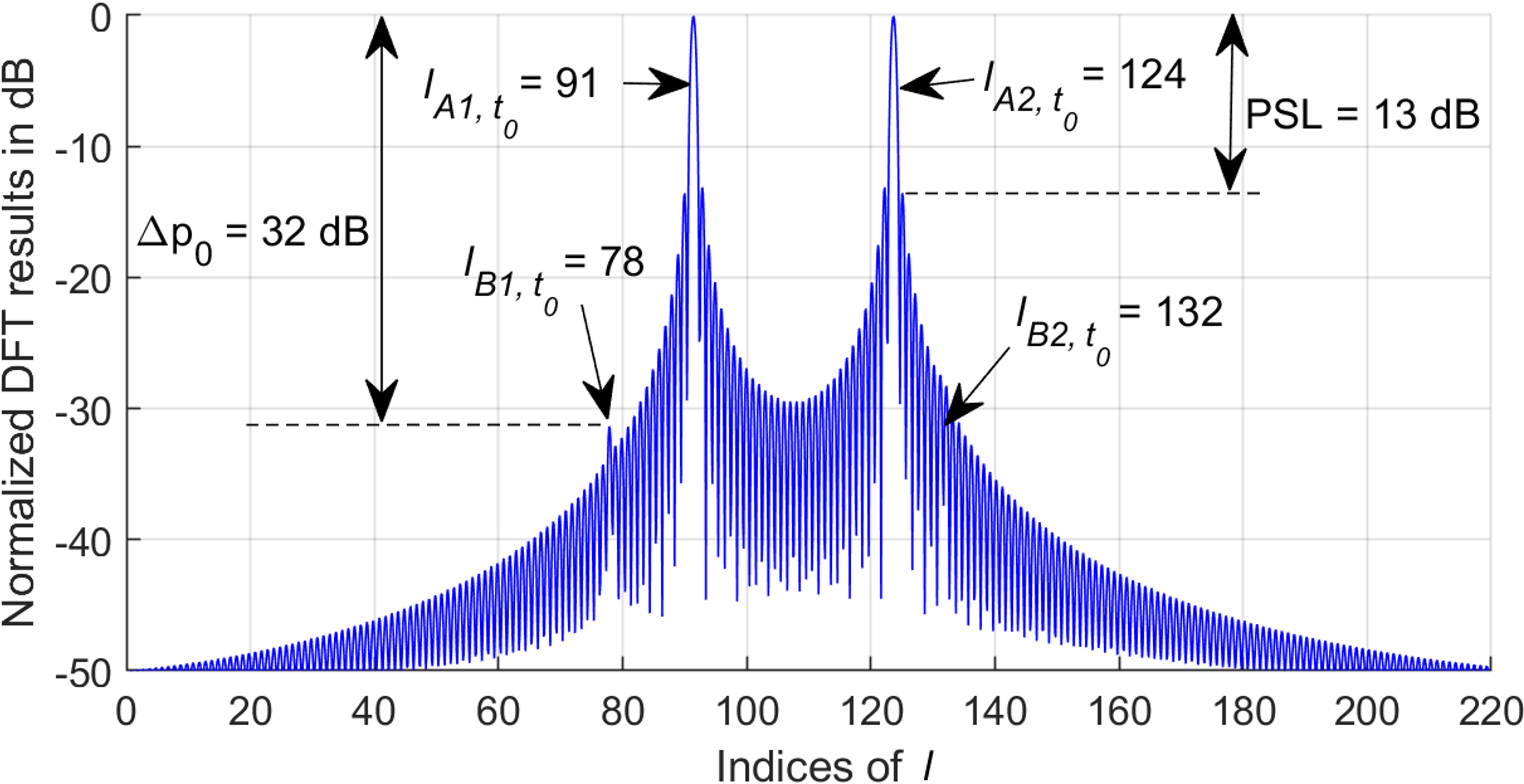}}\quad
	\subfloat[The radar image calculated at time $t_1$ = 0.2~s, vehicle-A with $R$ = 10~m and $v$ = 20~m/s, and vehicle-B with $R$ = 40~m and $v$ = 5~m/s, no windowing]{\label{fig:i}\includegraphics[width=0.65\columnwidth]{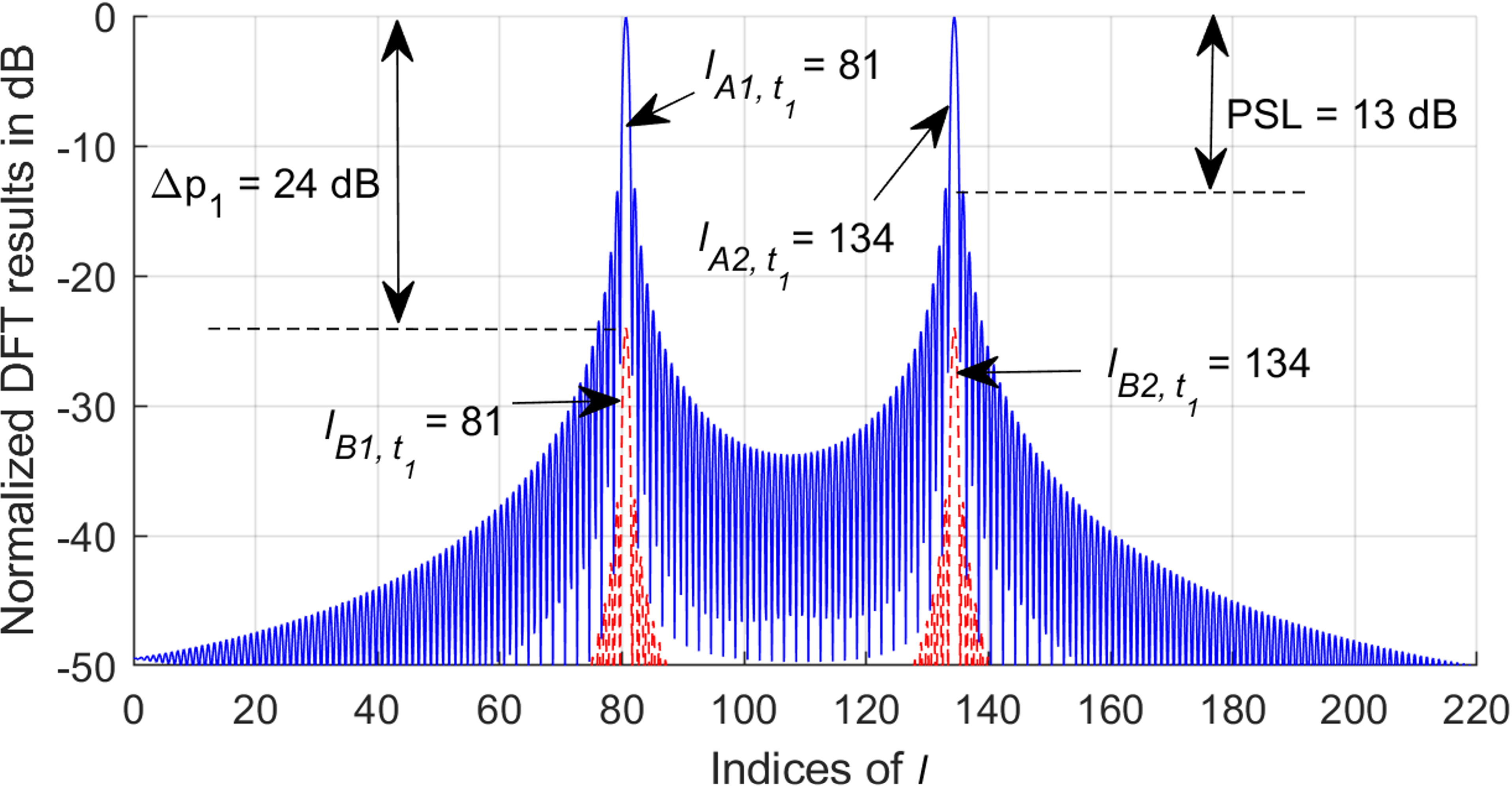}}\quad
	\subfloat[The radar image calculated at time $t_2$ = 0.6~s, vehicle-A with $R$ = 18~m and $v$ = 20~m/s, and vehicle-B with $R$ = 42~m and $v$ = 5~m/s, no windowing]{\label{fig:g}\includegraphics[width=0.65\columnwidth]{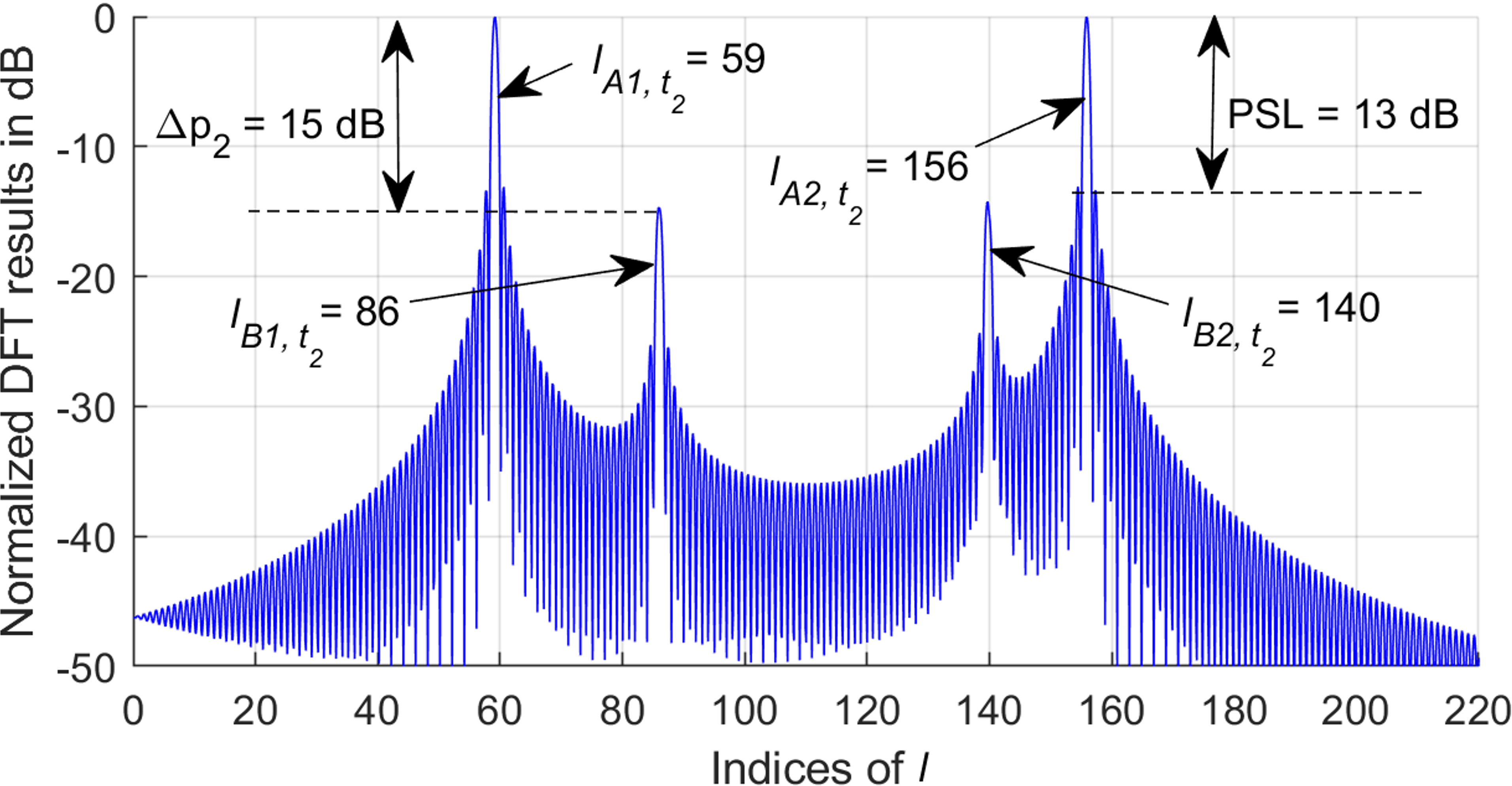}}\quad
	\subfloat[The radar image calculated at time $t_0$ = 0~s, vehicle-A with $R$ = 6~m and $v$ = 20~m/s, and vehicle-B with $R$ = 39~m and $v$ = 5~m/s, with Hamming window applied]{\label{fig:h}\includegraphics[width=0.65\columnwidth]{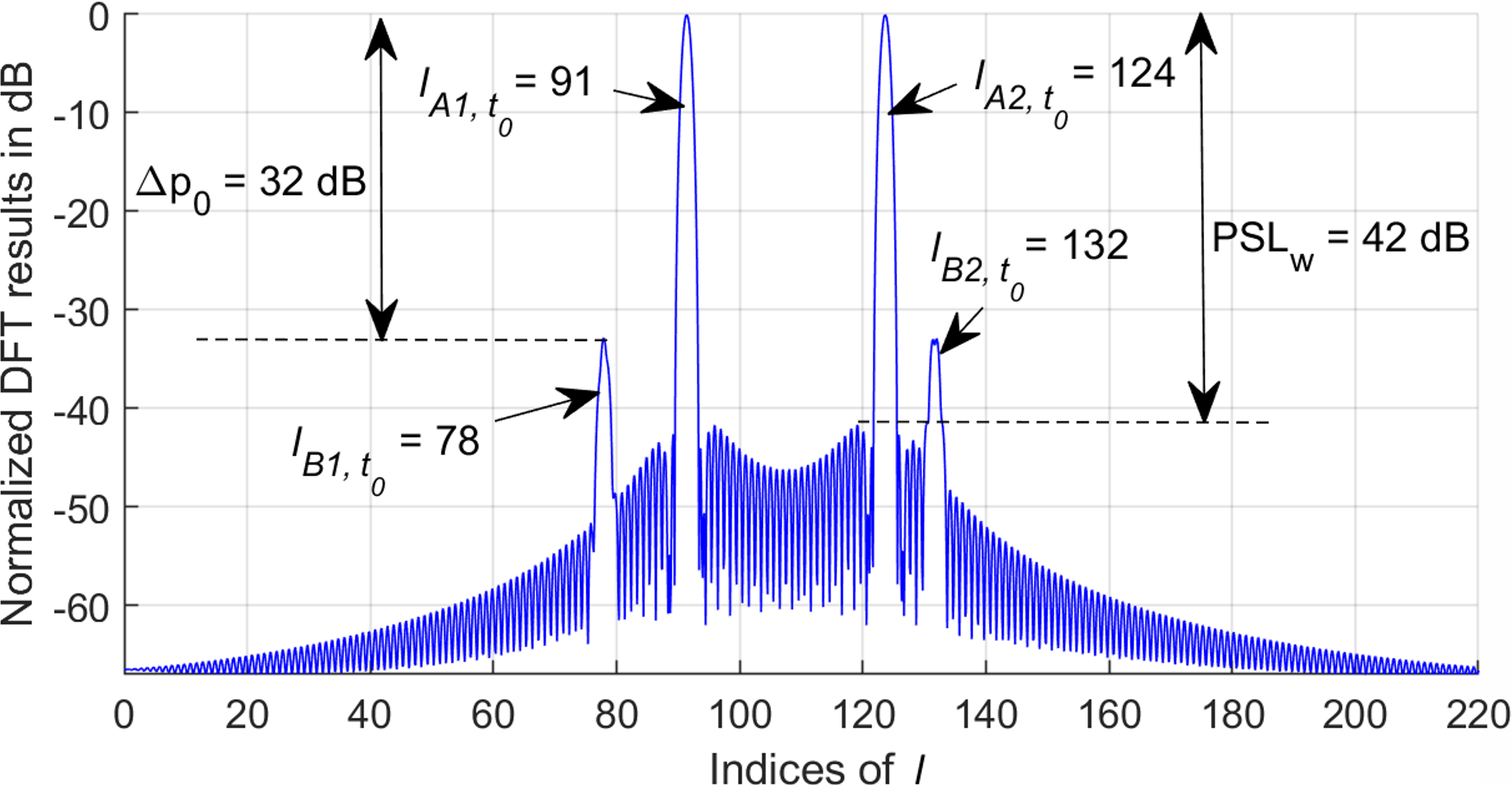}}\quad
	\subfloat[The radar image calculated at time $t_1$ = 0.2~s, vehicle-A with $R$ = 10~m and $v$ = 20~m/s, and vehicle-B with $R$ = 40~m and $v$ = 5~m/s, with Hamming window applied]{\label{fig:i}\includegraphics[width=0.65\columnwidth]{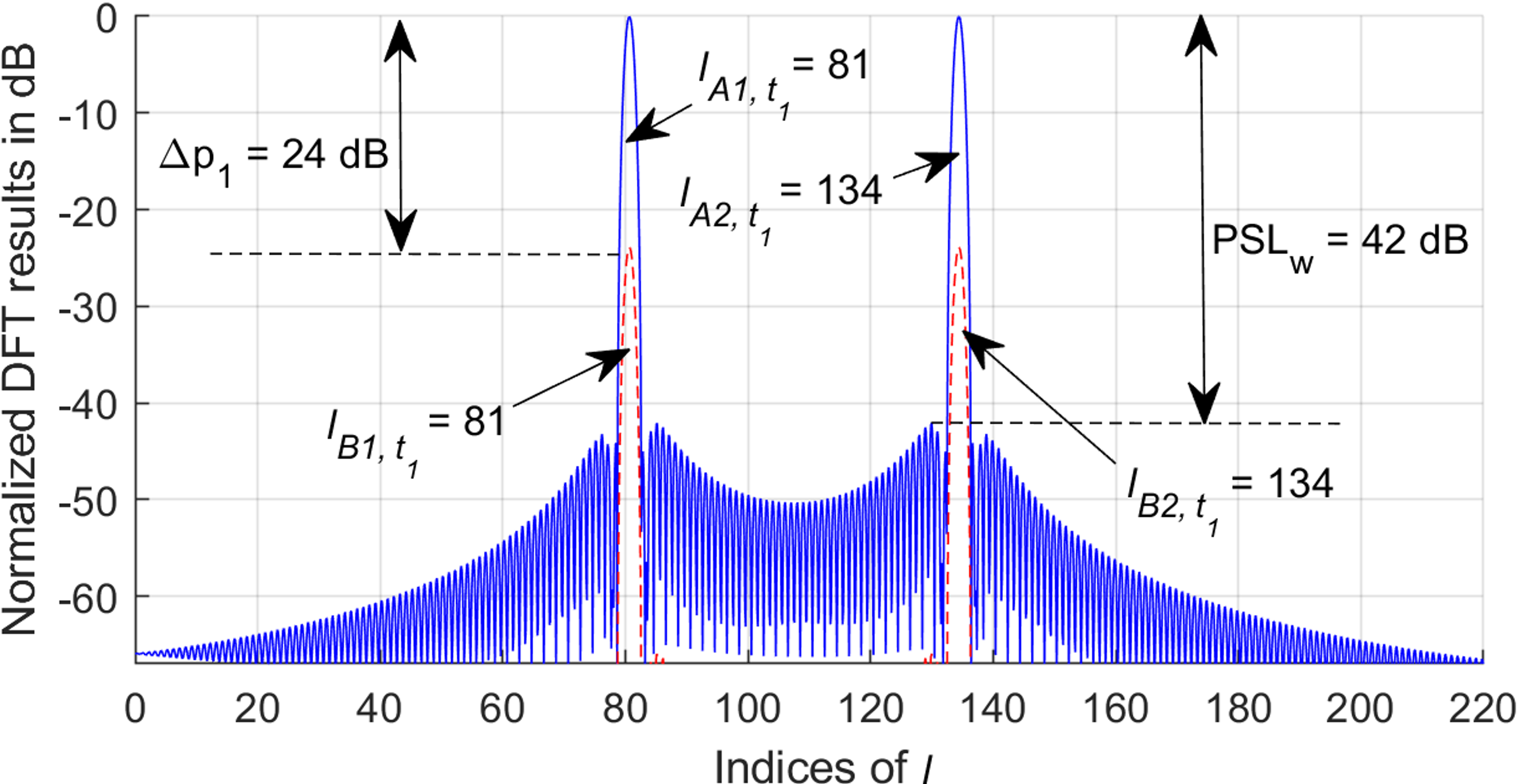}}\quad
	\subfloat[The radar image calculated at time $t_2$ = 0.6~s, vehicle-A with $R$ = 18~m and $v$ = 20~m/s, and vehicle-B with $R$ = 42~m and $v$ = 5~m/s, with Hamming window applied]{\label{fig:g}\includegraphics[width=0.65\columnwidth]{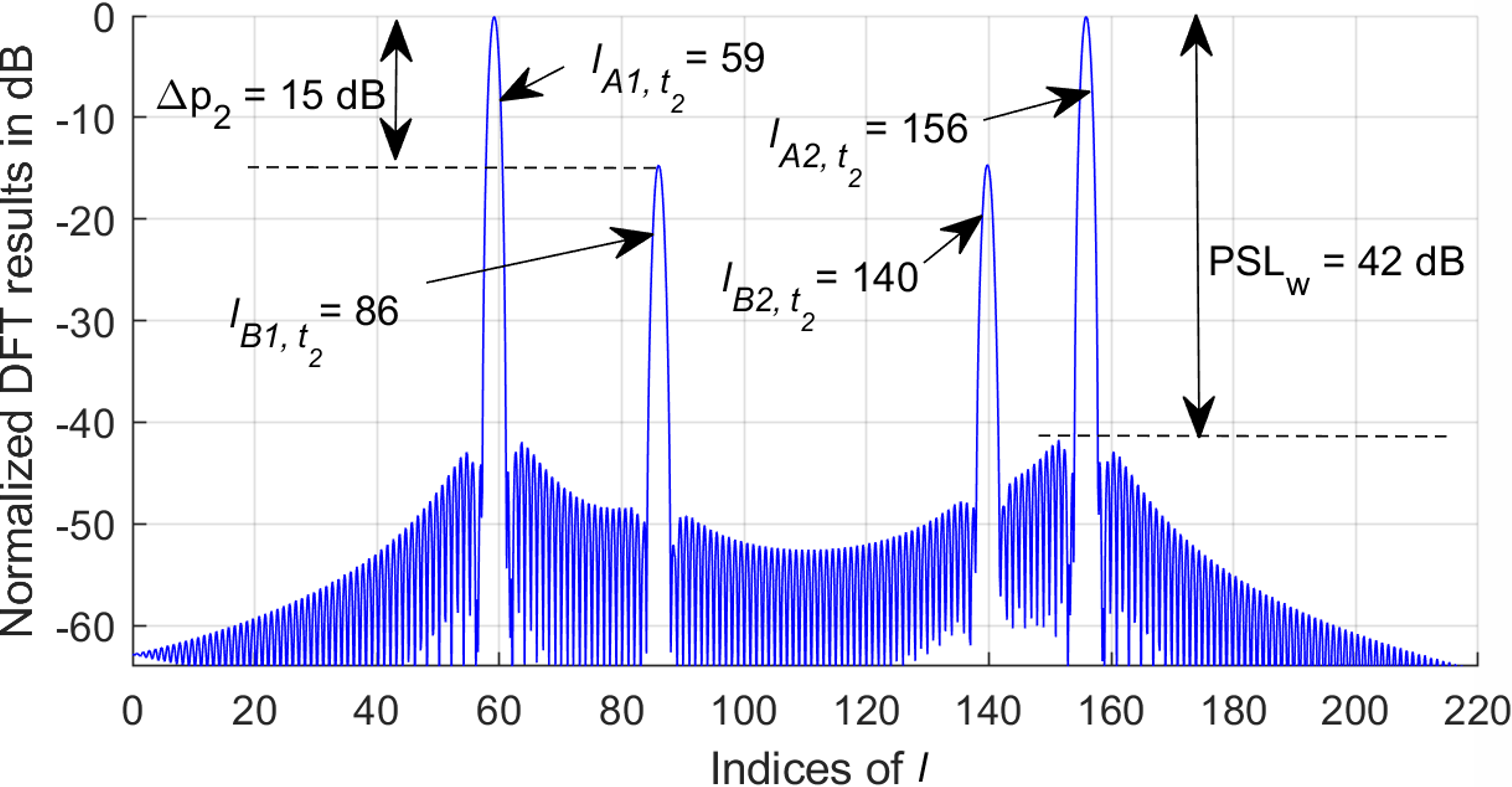}}\\	
	\caption{A simulated highway scenario and the radar images obtained by the ego vehicle using the proposed algorithm.}
	\label{fig_4}
\end{figure*}

\section{Performance Analysis for Multi-Object Scenario}
Our previous work in \cite{10107516} used a single target to evaluate the performance of the diagonal algorithm but did not verify its performance in multi-object scenarios. However, the typical scenarios, e.g., the automotive environment, consists of multiple objects. When applying the diagonal scheme to the multi-object scenario, additional problems arise, including:

\begin{itemize}
\item In the radar image, the amplitudes of the peaks induced by objects are determined by their radar cross sections (RCSs) and distances. The peak caused by a weak target (e.g., a tiny target or a target at a long distance) might be overshadowed by the sidelobes caused by a strong target (e.g., a large target or a target at close range). This problem reduces the probability of detection for weak targets.
\item Using the diagonal algorithm, two objects with different ranges and velocities may induce the same dual-peak profile. Unlabeled two peaks from a dual-peak profile in the radar image cause this ambiguity.
\item Each range-velocity estimate is derived from the two peaks caused by one object. The peaks in a radar image must be paired before applying the diagonal algorithm in multi-object scenarios.
\end{itemize}

To present our solutions for the above problems, we simulate a highway scenario with multiple vehicles and generate detections using the diagonal algorithm. As shown in Fig.~4, the scenario consists of a blue ego car with sensor and two cars to be detected, colored orange and red, respectively. All vehicles have identical RCS $\sigma = 5~\text{dBm}^2$. The ego car travels along the highway's center lane at a constant velocity of 20~m/s. At time \mbox{$t_0$ = 0~seconds}, the orange car, which is labeled as vehicle-A, is moving 6~meters ahead of the ego vehicle at a constant velocity of 40~m/s along the left lane, while the red car (vehicle-B) is traveling 39~m in front of the ego car at a constant velocity of 25~m/s along the right lane. \mbox{Figs.~4(a)-(c)} illustrate the snapshots of this highway scenario captured at time \mbox{$t_0$ = 0~seconds}, \mbox{$t_1$ = 0.2~seconds}, and \mbox{$t_2$ = 0.6~seconds}, respectively, showing the situation in front of the ego car using chase camera view and top view.

\begin{figure*}[t]
	\centering
	\subfloat[Snapshot of a simulated highway scenario showing a yellow car, a purple motorcycle, and a red truck in front of the blue ego car]{\label{fig:h}\includegraphics[width=0.65\columnwidth]{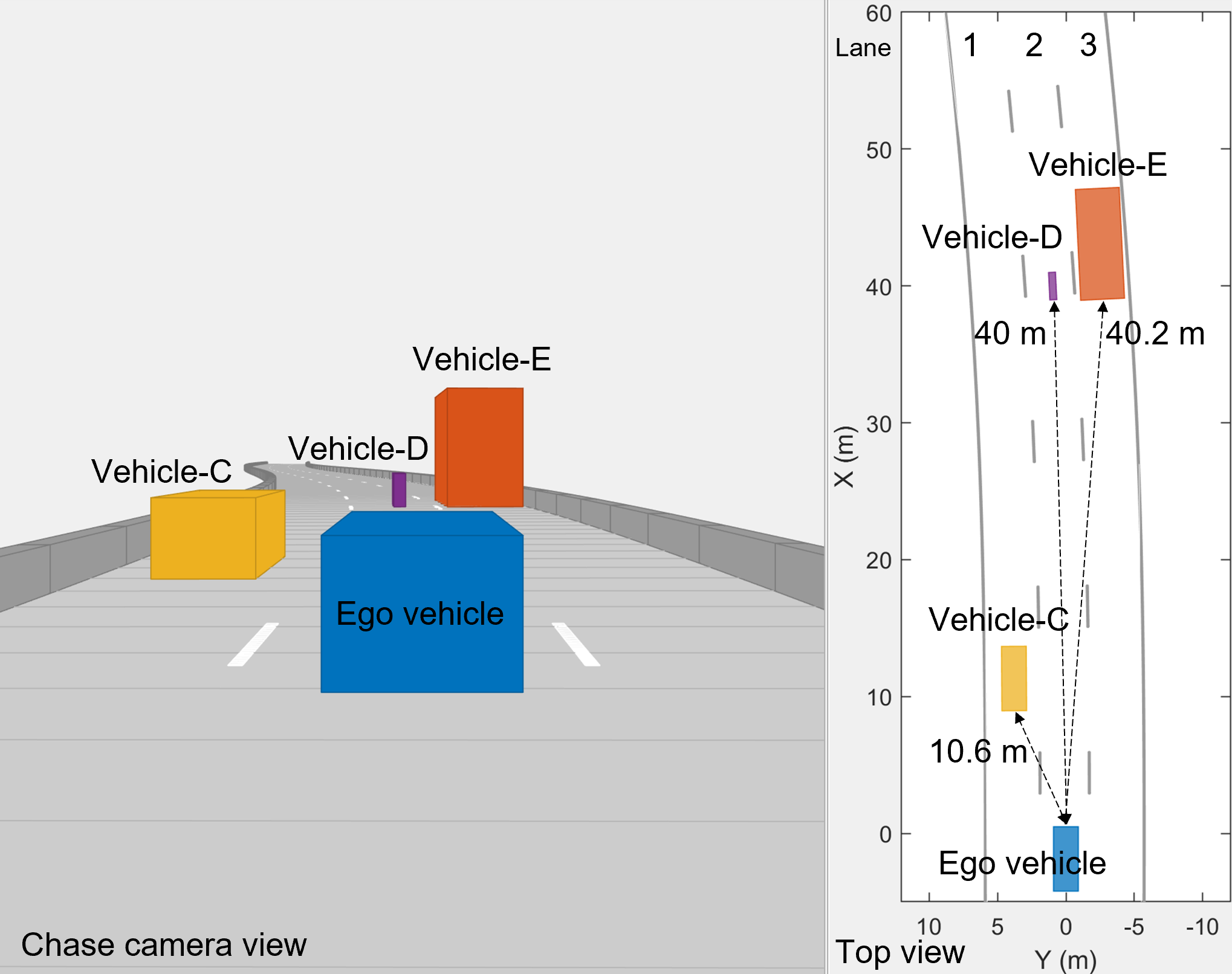}}\quad
	\subfloat[The radar image without window function, vehicle-C and vehicle-E can be identified, but vehicle-D is undetectable ]{\label{fig:i}\includegraphics[width=0.65\columnwidth]{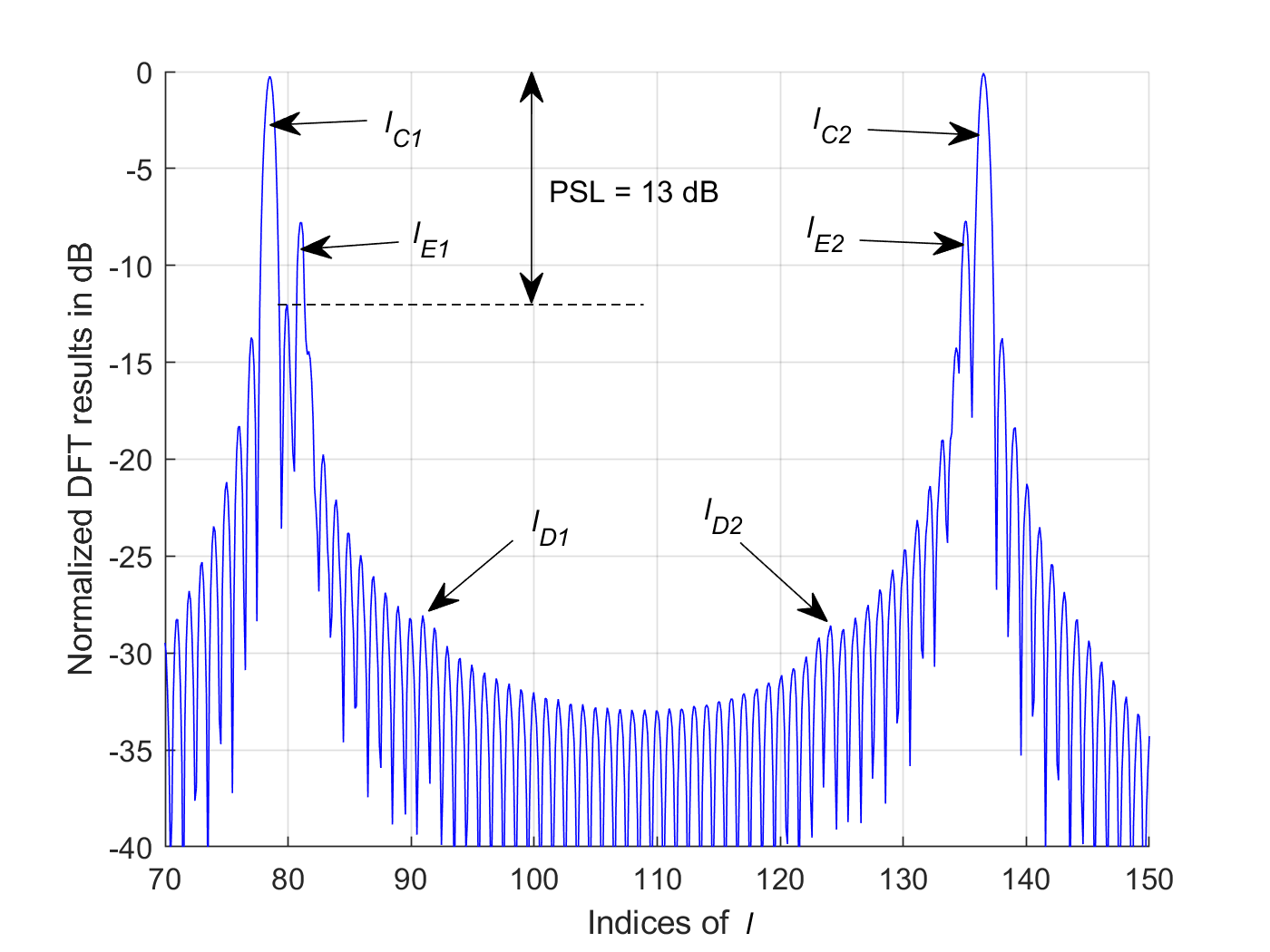}}\quad
	\subfloat[The radar image with Hamming window, vehicle-D is detectable, but vehicle-C and \mbox{vehicle-E} cannot be separated ]{\label{fig:g}\includegraphics[width=0.65\columnwidth]{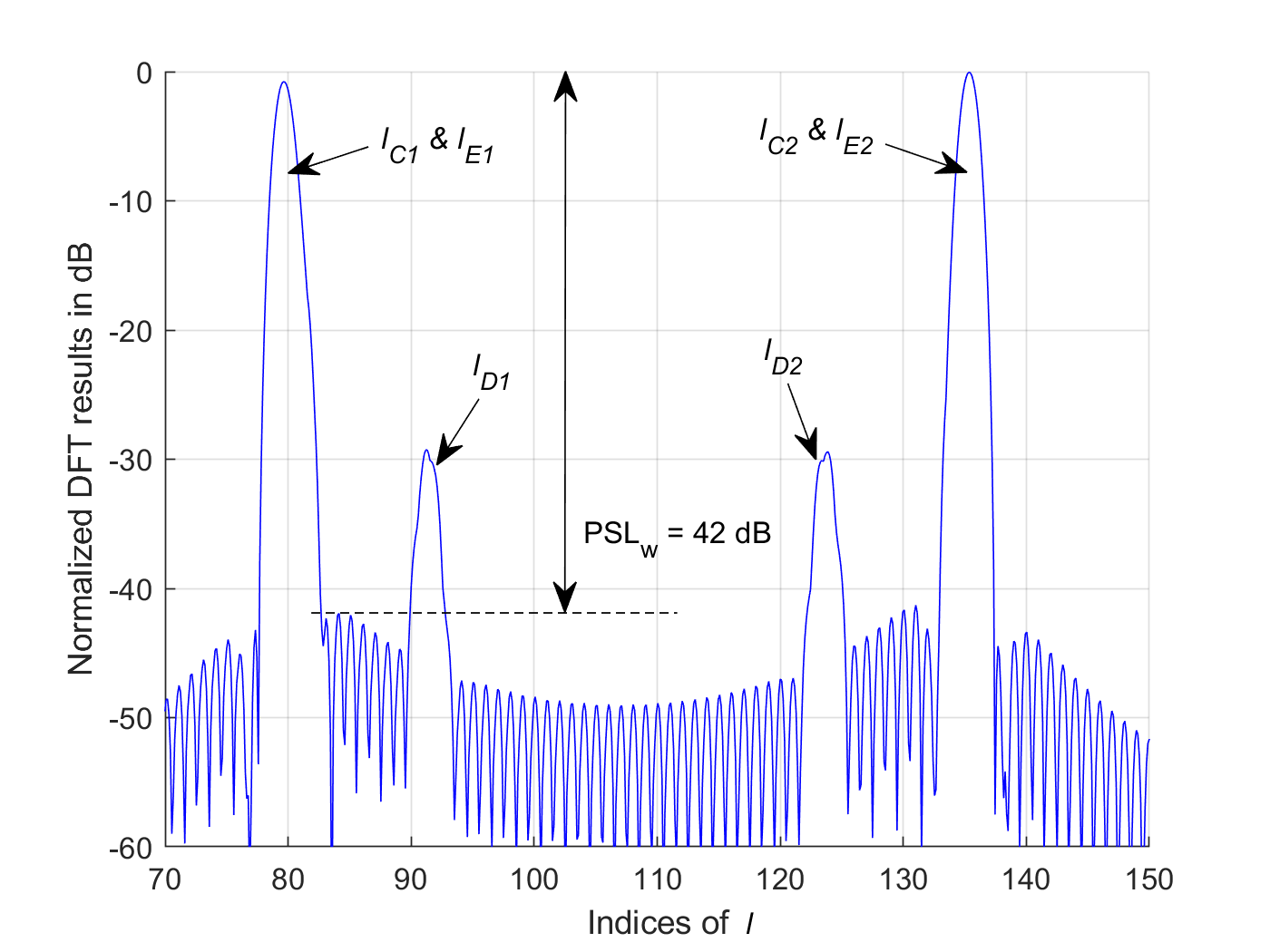}}\\
	\caption{A simulated highway scenario and the radar images obtained by the ego vehicle using the proposed algorithm.}
	\label{fig_5}
\end{figure*}

The ego car is equipped with a forward-looking antenna that simultaneously provides communication and range-velocity sensing coverage. Angle estimation is not considered in this paper. Fig.~4(d) is derived from the measurement at time \mbox{$t_0$ = 0~seconds} using the diagonal algorithm. The two dual-peak profiles resulting from vehicle-A and vehicle-B are superimposed. However, only the two peaks, $l_{\text{A1},t_0}$ and $l_{\text{A2},t_0}$, caused by vehicle-A, are sharply depicted. The peaks caused by vehicle-B ($l_{\text{B1},t_0}$ and $l_{\text{B2},t_0}$) are overshadowed by the sidelobes caused by vehicle-A. The amplitude of the peak in the radar image relates to the received signal power $P_\text{R}$ reflected by the object. $P_\text{R}$ is given by \cite{6140075}
\begin{equation}\label{eqn_17}
	P_\text{R} = \frac{P_{\text{Tx}}G_{\text{Tx}}G_{\text{Rx}}\sigma\lambda^2}{(4\pi)^3R^4f_\text{c}^2},
\end{equation}
where $P_{\text{Tx}}$ is the transmitted power, $G_{\text{Tx}}$ is the antenna gain at Tx, $G_{\text{Rx}}$ is the antenna gain at Rx, $\sigma$ is the RCS of the object, $\lambda$ is the wavelength of the carrier, $R$ is the distance between the antenna and the object.

From \eqref{eqn_17}, the received signal power $P_\text{R}$ is dependent on range $R$. The return signal from a close target is stronger than the return signal from a target farther away if they have identical RCS. According to their distances, the received signal power from \mbox{vehicle-A} is 32~dB higher than the received signal power from vehicle-B. This results in the fact that the amplitudes of peak $l_{\text{A1},t_0}$ and $l_{\text{A2},t_0}$ caused by \mbox{vehicle-A} are 32~dB higher than the amplitudes of peaks $l_{\text{B1},t_0}$ and $l_{\text{B2},t_0}$ caused by vehicle-B. Peak $l_{\text{B2},t_0}$ is totally overshadowed by the sidelobes caused by vehicle-A, and peak $l_{\text{B1},t_0}$ is almost indistinguishable. Therefore, vehicle-B cannot be detected. This exhibits that the dynamic range of the JCAS system is limited by the sidelobes of the Fourier transform when close targets or large targets exist in the detection area.

In the discrete-time signal process, Hamming windowing is a signal shaping technique, usually applied to mitigate the impact of sidelobes of DFT \cite{1455106}. Fig.~4(g) shows the radar image derived from the scenario illustrated in Fig.~4(a) with the diagonal algorithm and with Hamming window. Compared to Fig.~4(d), the peak-to-sidelobe level (PSL) with Hamming window $\text{PSL}_\text{w}$ is reduced from \mbox{-13 dB} to \mbox{-42~dB}. With the help of Hamming window, peaks $l_{\text{B1},t_0}$ and $l_{\text{B2},t_0}$ caused by vehicle-B now clearly stand out against the well-suppressed sidelobes induced by vehicle-A.

Fig.~4(e) and Fig.~4(h) illustrate the radar images derived by the ego car at time \mbox{$t_1$ = 0.2~seconds} without and with Hamming window, respectively. A red dashed line indicates the dual-peak profile caused by vehicle-B for clear visualization. As displayed, the bin positions of peaks $l_{\text{A1},t_1}$ and $l_{\text{A2},t_1}$ coincide with the bin positions of peaks $l_{\text{B1},t_1}$ and $l_{\text{B2},t_1}$. This is the ambiguity mentioned in Section~III, i.e., two objects with different ranges and velocities may induce the same dual-peak profile. Range and velocity of vehicle-A and vehicle-B cannot be determined only by the superimposed dual-peak profiles derived at time $t_1$. However, variations in range and velocity cause time-varying peak amplitude and peak bin index in the radar image. Therefore, the fusion of estimates acquired at various times can resolve the ambiguity. For example, according to the different peak patterns of vehicle-A and vehicle-B at time $t_0$, the range-velocity estimates derived at time $t_0$ can assist the JCAS system in estimating the range-velocity of vehicle-A and vehicle-B when ambiguity occurs at time $t_1$. How to resolve this ambiguity is analyzed in greater detail in our previous work \cite{10107516}.

As illustrated in \mbox{Figs.~4(a)-(c)}, vehicle-A is approaching vehicle-B due to its higher velocity than the velocity of vehicle-B. Therefore, the difference between the received signal powers from both vehicles is decreasing. As shown in Fig.~4(f), at time \mbox{$t_2$ = 0.6~seconds}, the peaks induced by vehicle-B ($l_{\text{B1},t_2}$ and $l_{\text{B2},t_2}$) stand out against the sidelobes caused by vehicle-A, even though Hamming window is not applied. This is because the sidelobes caused by vehicle-A are not higher than peaks $l_{\text{B1},t_2}$ and $l_{\text{B2},t_2}$ any longer due to the similar signal returns from both vehicles. The amplitude difference between peaks caused by the two vehicles $\Delta{p_2}$ reduces to 15~dB only.

From Figs.~4(d)-(i), it can be concluded that weak targets might be overshadowed when strong targets are in the detection range. A window with high sidelobe attenuation can be used to mitigate this issue. However, it reduces the ``peak resolution" of the diagonal algorithm. It is worth mentioning that peak resolution here differs from the radar's definition of range resolution (or velocity resolution). The bin index of a peak obtained by the diagonal algorithm is mixed with the range and velocity information of the object. Hence, two objects with vastly different ranges (or velocities) may induce peaks close to each other. Peak resolution is the capability of the diagonal algorithm to distinguish adjacent peaks caused by two objects in the radar image.

We investigate the performance of the diagonal algorithm with and without applying Hamming window in a simulated highway scenario shown in Fig.~5(a). The scenario consists of two strong scatterers and one weak scatterer, which are an orange car (vehicle-C, strong scatterer) with a small RCS of 5 $\text{dBm}^2$ close to the ego car, a purple motorcycle (vehicle-D, weak scatterer) with a small RCS of 1 $\text{dBm}^2$ far away from the ego car, and a red truck (vehicle-E, strong scatterer) with a large RCS of 100 $\text{dBm}^2$ far away from the ego car. \mbox{Vehicle-C} is moving 10.6~m ahead of the blue ego car at a relative velocity of 20~m/s on the left lane. Vehicle-D is traveling 40~m in front of the ego car on the same lane at a relative velocity of 3~m/s. Vehicle-E, 40.2~m away from the ego car, is moving at a relative velocity of 5~m/s on the right lane.

Fig.~5(b) shows the radar image derived by the ego vehicle using the diagonal algorithm. Vehicle-C causes peaks $l_\text{C1}$ and $l_\text{C2}$, and vehicle-E causes peaks $l_\text{E1}$ and $l_\text{E2}$. Although the truck is far away from the ego car, the return signal from the truck is only 8~dB weaker than the return signal from the strongest scatterer (vehicle-C) thanks to the large RCS of the truck. Peaks $l_\text{E1}$ and $l_\text{E2}$ can still stand out against the unsuppressed sidelobes caused by \mbox{vehicle-C}. \mbox{Vehicle-C} and \mbox{vehicle-E} cause peaks close to each other. However, the diagonal algorithm without window function has high peak resolution due to the narrow mainlobes in the radar image. The peaks caused by \mbox{vehicle-C} and \mbox{vehicle-E} can be clearly separated. The weak scatterer (vehicle-D) has a small RCS and is far from the radar. Thus, the return signal from the motorcycle is 30~dB weaker than that from vehicle-C. The peaks caused by the motorcycle, $l_\text{D1}$ and $l_\text{D2}$, are almost indistinguishable in the radar image.

Fig.~5(c) illustrates the radar image produced by the diagonal algorithm with Hamming window. One can see that the sidelobes of peaks caused by the strong scatterer are well suppressed but at the cost of lower peak resolution. The peaks caused by the weak scatterer, $l_\text{D1}$ and $l_\text{D2}$, can be detected. However, as a result of Hamming window, the peaks caused by vehicle-C and vehicle-E are widened and they cannot be separated in the radar image.

We suggest an adaptive window strategy, e.g., a time multiplexing window strategy, to ensure both the peak resolution and detection of weak objects when designing a JCAS system. Window function needs to be applied according to the distribution of the objects in the detection range. Peak resolution is essential if the radar image contains two or more peaks near to each other and the peaks have similar amplitudes. In this case, it is best to sense the objects without applying any window function. A window function should be used if there is a large gap between the return signal power from different objects.

Another problem with the diagonal algorithm is peak pairing. The algorithm uses the pairwise peaks caused by one object to derive its range and velocity. However, the dual-peak profiles caused by multiple objects are superimposed, producing a radar image with more peaks than the number of objects. Consequently, the diagonal algorithm cannot be directly applied. For example, as shown in Fig.~4(i), vehicle-A and vehicle-B induce four peaks in the radar image. The four peaks must be paired such that the peaks in each pair are caused by one object before applying the diagonal algorithm. A simple solution to this problem is to compare the amplitudes of all peaks in the radar image. According to \eqref{eqn_17}, the two peaks caused by one object have the same amplitude. Thus, it can be concluded that one vehicle causes peaks $l_{\text{A1},t_2}$ and $l_{\text{A2},t_2}$, and another vehicle causes peaks $l_{\text{B1},t_2}$ and $l_{\text{B2},t_2}$.

\section{Conclusions}
A diagonal waveform and corresponding 1D-DFT diagonal algorithm for the JCAS system was proposed in \cite{10107516}. Compared to the grid waveform structure in Fig.~1, the overhead of the diagonal waveform in Fig.~2 reduces from $N^2/N^2_c$ to $N/N^2_c$. Furthermore, the computational complexity of one estimation is reduced from $2N\times{\mathcal{O}(N^2)}$ to $\mathcal{O}(N^2)$. However, the diagonal algorithm in \cite{10107516} was only verified for single object scenarios. In this paper, We simulate a highway scenario including multiple vehicles with different ranges, velocities, and RCSs. We then demonstrate that the diagonal algorithm can be extended to range-velocity estimation for multi-object scenarios. A new concept ``peak resolution'' and some design strategies for the diagonal algorithm are also presented.

\section{Acknowledgment}
This work has been funded by the National Key Research and Development Program of China (No.2021YFB2900200).

\bibliographystyle{IEEEtran}
\bibliography{ref} 

\vspace{12pt}
\color{red}
\end{document}